\documentclass[onecolumn,aps,prc,showpacs,superscriptaddress]{revtex4}
\usepackage{epsfig}

\newcommand {\pt}	{p_{T}}
\newcommand {\phit}	{\phi_{t}}

\newcommand {\psiRP}	{\psi_{RP}}
\newcommand {\dphi}	{\Delta\phi}
\newcommand {\cij}	{\cos2\dphi_{ij}}
\newcommand {\cphi}	{\cos2\dphi}
\newcommand {\sphi}	{\sin2\dphi}
\newcommand {\tphi}	{\tilde\phi}
\newcommand {\intpi}	{\int_{0}^{2\pi}}
\newcommand {\vtwo}[1]	{v_{2,#1}}
\newcommand {\vv}[1]	{v_2\{#1\}}
\newcommand {\vsq}[1]	{v_2^2\{#1\}}
\newcommand {\mean}[1]	{\langle#1\rangle}
\newcommand {\acl}	{{a}}
\newcommand {\Phy}	{\mathcal{P}_{hy}}
\newcommand {\Pa}	{\mathcal{P}_{\acl}}
\newcommand {\Nhy}	{N_{hy}}
\newcommand {\Ncl}	{N_{cl}}
\newcommand {\Na}	{N_{\acl}}
\newcommand {\Nt}	{N_{t}}
\newcommand {\ns}	{{ns}}
\newcommand {\as}	{{as}}
\newcommand {\Nbg}	{N_{bg}}
\newcommand {\Njet}	{N_{a,jet}}
\newcommand {\Nak}	{N_{a,k}}
\newcommand {\fcl}	{f_{\acl}}
\newcommand {\rhocl}	{\rho_{cl}}
\newcommand {\be}	{\begin{equation}}
\newcommand {\ee}	{\end{equation}}
\newcommand {\bea}	{\begin{eqnarray}}
\newcommand {\eea}	{\end{eqnarray}}

\begin{document}

\title{Non-flow, and what flow to subtract in jet-correlation}

%\ShortTitle{Non-flow, and what flow to subtract in jet-correlation}

%\author{\speaker{Fuqiang Wang}\thanks{This work is supported by U.S. Department of Energy under Grant DE-FG02-88ER40412.}\\
%\author{\speaker{Fuqiang Wang}\\
%  Purdue University, 525 Northwestern Avenue, West Lafayette, Indiana, USA\\
%  E-mail: \email{fqwang@purdue.edu}}

%\author{Quan Wang\\
%  Purdue University, 525 Northwestern Avenue, West Lafayette, Indiana, USA\\
%  E-mail: \email{wang187@purdue.edu}}
\author{Quan Wang}
\author{Fuqiang Wang}
\affiliation{Department of Physics, Purdue University, 525 Northwestern Ave., West Lafayette, IN 47907}

\begin{abstract}
We derive analytical forms for non-flow contributions from cluster correlation to two-particle elliptic flow ($\vv{2}$) measure. 
  %We derive an analytical form for jet-correlation flow-background in a cluster approach. 
  We also derive an analytical form for jet-correlation flow-background with the same cluster approach.
  We argue that the elliptic flow $v_2$ parameter to be used in jet-correlation background is that from two-particle method excluding non-flow correlations unrelated to the reaction plane, but including cross-terms between cluster correlation and cluster flow. We verify our result with Monte Carlo simulations. We discuss how one may obtain the $v_2$ parameter for jet-correlation background experimentally. %We examine jet-correlation data from STAR and PHENIX using our refined flow-background. We discuss implications of our result on Mach-cone signal and jet modification in general.
\end{abstract}

%\FullConference{High-pT Physics at LHC -09\\
%		 February 4- 4 2009\\
%		 Prague, Czech Republic}

\maketitle

\section{Introduction}

Jet-like angular correlation studies with high transverse momentum ($\pt$) trigger particles have provided valuable information on the properties of the medium created in relativistic heavy-ion collisions~\cite{STARwp,PHENIXwp}. In such studies, correlation functions are formed in azimuthal angle difference between an associated particle and a high $\pt$ trigger particle, which preferentially selects (di-)jet. One important aspect of these studies is the subtraction of combinatorial background which itself is non-uniform due to anisotropic particle distribution with respect to the reaction plane-- both the trigger particle and the associated particles are correlated with the common reaction plane in an event. The critical part is to determine flow parameters, mainly elliptic flow ($v_2$), to be used in constructing background. 

There are many $v_2$ measurements~\cite{v4,v2data}. They contain various degrees of non-flow contributions, such as those from resonance decays and jet correlations. %Some of those non-flow effects should be included in jet-correlation background. For simplicity, we shall still refer to this background generally as flow-background. 
Those non-flow effects should not be included in jet-correlation background. 
We shall refer to this jet-correlation background as flow-background. 
The anisotropic flow to be used for flow-background should be ideally that from two-particle method, $\vv{2}$~\cite{v2method,v2method2}, because jet-like correlation is analyzed by two-particle correlation method. Moreover, two-particle anisotropic flow contains fluctuations which should be included in jet-correlation flow-background~\cite{v2method,v2method2}. 

Non-flow is due to azimuthal correlations unrelated to the reaction plane, such as resonances, (mini)jets, or generally, clusters. In this proceedings, we study non-flow contributions in two-particle $\vv{2}$ in a cluster approach as in~\cite{FWang}; analytical form is derived for each non-flow component. We shall demonstrate that the flow to be used in jet-correlation background subtraction should be the two-particle $\vv{2}$ excluding cluster correlations unrelated to the reaction plane, but including cross-terms between cluster correlation and cluster flow~\cite{FWang2}. We verify our analytical result with Monte Carlo simulations. 
We discuss how one may obtain the elliptic flow $v_2$ parameter for jet-correlation background experimentally.
%We then examine available jet-correlation data from both STAR~\cite{STARwp} and PHENIX~\cite{PHENIXwp} with our refined flow-background. Finally we discuss implications of our result on the observed Mach-cone signal~\cite{3part} and on medium modification of jets in general.

\section{Non-flow effect from cluster correlations\label{sec:nonflow}}

Suppose an event is composed of particles from hydro-medium and clusters of various types (such as minijets and resonance decays). Particle pairs can be decomposed into four sources: 
\begin{itemize}
\item particle pairs from hydro-medium ($B$),
\item particle pairs from same cluster ($C$),
\item particle pairs between hydro-medium and clusters ($X$), and
\item particle pairs between clusters ($Y$).
\end{itemize}
The total sum of the cosines of pair opening angles is
\be
\sum_{i\neq j}\cij=B+\sum_{k\in cluster}C+\sum_{k\in cluster}2X+\sum_{(k_1\neq k_2)\in cluster}Y,
\label{eq1}
\ee
where
\bea
B&=&\sum_{(i\neq j)\in hydro}\cij,\\
C&=&\sum_{(i\neq j)\in k}\cij,\\
X&=&\sum_{i\in k}\sum_{j\in hydro}\cij,\\
Y&=&\sum_{i\in k_1}\sum_{j\in k_2}\cij.
\eea
Here $i,j$ are particle indices, $\dphi_{ij}=\phi_i-\phi_j$, and $k$ stands for a cluster. Below we derive analytical form for each source.

\subsection{Background flow correlation}

Hydro-background particle correlation is only from hydrodynamic anisotropic flow:
\be
B=\sum_{(i\neq j)\in hydro}\cij=\Phy\mean{\cphi_{hy}}=\Phy\vsq{2}_{hy}
\ee
where $\Phy=\mean{\Nhy(\Nhy-1)}$ is the number of background pairs.

\subsection{Particle correlation within cluster}

Particle correlation within cluster is given by
\bea
C=\sum_{(i\neq j)\in k}\cij&=&\intpi\Pa(\tphi_k)\rhocl(\tphi_k)d\tphi_k\times\nonumber\\
&&\intpi \fcl(\dphi_i,\tphi_k)\dphi_i\intpi \fcl(\dphi_j,\tphi_k)d\dphi_j\cos2(\dphi_i-\dphi_j).
\eea
Here $\dphi_{i,j}=\phi_{i,j}-\phi_k$ is the azimuthal angles of particles in cluster $k$ relative to cluster axis $\phi_k$ (which can be defined just for convenience); $\fcl(\dphi,\tphi_k)$ is the correlation function of (associated) particles inside cluster $k$ relative to the cluster axis $\phi_k$, generally dependent of the cluster axis $\tphi_k=\phi_k-\psiRP$ relative to the reaction plane, and $\intpi \fcl(\dphi,\tphi_k)\dphi\equiv1$; $\Pa(\tphi_k)$ is number of (associated) particle pairs in cluster $k$, generally dependent of the cluster axis; $\rhocl(\tphi_k)$ is the density function of cluster $k$ relative to the reaction plane, which we will assume is given by elliptic flow of clusters, and $\intpi\rhocl(\tphi_k)d\tphi_k\equiv1$. Note, there can be many types of clusters (e.g. jet-correlation, resonance decays); the subscript `$cl$' stands for one type of clusters and we have omitted summation over all types of clusters from the formulism; in this work we will discuss only one type of clusters at a time. 

In general,
\be
C=\Pa\mean{\cij}_{cl}
\ee
where $\Pa$ is average number of pairs per cluster and $\mean{\cij}_{cl}$ is the average cosine of twice pair opening angle in the cluster. 

If particles inside cluster are independent of each other except all of them are correlated with the cluster axis, then we can factorize the correlation terms and obtain
\bea
C&=&\intpi\Pa(\tphi_k)\rhocl(\tphi_k)d\tphi_k\left[\left(\intpi \fcl(\dphi,\tphi_k)\cphi d\dphi\right)^2+\left(\intpi \fcl(\dphi,\tphi_k)\sphi d\dphi\right)^2\right]\nonumber\\
&=&\intpi\Pa(\tphi_k)\left(\mean{\cphi}_{\tphi_k}^2+\mean{\sphi}_{\tphi_k}^2\right)\rhocl(\tphi_k)d\tphi_k.
\eea
Here $\mean{\cphi}_{\tphi_k}$ and $\mean{\sphi}_{\tphi_k}$ $(\dphi=\phi-\phi_k)$ are averages within cluster $k$, and are generally dependent of the cluster axis $\tphi_k$. 

If the cluster correlation function $\fcl(\dphi,\tphi_k)$ is symmetric about $\dphi=0$, then $\mean{\sphi}_{\tphi_K}=0$. Further, in the special case where particle correlation in clusters does not vary with cluster location $\tphi_k$, i.e., $\Pa={\rm const.}$ and $\fcl(\dphi,\tphi_k)={\rm const.}$, then
\be
C=\Pa\mean{\cphi}_{cl}^2.
\ee

\subsection{Background-cluster correlation}

Correlation between cluster particles and hydro-medium particles is given by
\bea
X=\sum_{i\in k}\sum_{j\in hydro}\cij&=&\intpi \Nhy(\tphi_{k=1,2,...})\rho_{hy}(\tphi_{hy})d\tphi_{hy}\intpi\rhocl(\tphi_k)d\tphi_k\times\nonumber\\
&&\intpi\Na(\tphi_k)\fcl(\dphi_i,\tphi_k)d\dphi_i\left[\cos2(\phi_i-\phi_{hy})\right].
\eea
Here $\tphi_{hy}=\phi_{hy}-\psiRP$, and $\rho_{hy}(\tphi_{hy})$ is the density function of hydro particles relative to the reaction plane (i.e., anisotropic hydro flow); $\Na(\tphi_k)$ is number of (associated) particles in cluster and is generally dependent of the cluster axis $\tphi_k$. For generality we have taken the number of hydro-medium particles $\Nhy(\tphi_{k=1,2...})$  to depend on positions of all clusters. Such dependence can arise in real data analysis, such as jet-correlation analysis, from interplays between centrality cut and biases due to selection of specific clusters. Rewriting $\phi_i-\phi_{hy}=\dphi_i+\tphi_k-\tphi_{hy}$, we have
\bea
X&=&\intpi \Nhy(\tphi_{k=1,2,...})\rho_{hy}(\tphi_{hy})\cos2\tphi_{hy}d\tphi_{hy}\times\nonumber\\
&&\intpi\rhocl(\tphi_k)d\tphi_k\intpi \Na(\tphi_k)\fcl(\dphi_i,\tphi_k)\cos2(\dphi_i+\tphi_k)d\dphi_i.
\eea
Here we have used $\displaystyle{\intpi}d\tphi_k\rhocl(\tphi_k)\displaystyle{\intpi}d\dphi_k\fcl(\dphi_k,\tphi_k)\sin2(\dphi_k+\tphi_k)=0$ because of symmetries $\fcl(\dphi_k,\tphi_k)=\fcl(-\dphi_k,-\tphi_k)$ and $\rhocl(\tphi_k)=\rhocl(-\tphi_k)$. Note, due to elliptic flow of clusters, cluster particles acquire elliptic flow
\be
\vtwo{\acl}\equiv\mean{\cos2(\phi-\psiRP)}=\frac{1}{\Na}\intpi\rhocl(\tphi_k)d\tphi_k\intpi \Na(\tphi_k)\fcl(\dphi_i,\tphi_k)\cos2(\dphi_i+\tphi_k)d\dphi_i.
\label{eq6}
\ee
Using the notation in Eq.~(\ref{eq6}), we have
\be
X=\Nhy\Na \vtwo{hy}\vtwo{\acl}=\Nhy\Na\vv{2}_{hy}\vv{2}_{\acl}.
\ee
Here the product of the $v_2$'s includes flow fluctuation, and equals to the product of two-particle $v_2$'s. This is because $\vv{2}$ of hydro-particles and cluster particles contain only fluctuation; non-flow does not exist between hydro-particles, nor between particles from different clusters. (Note, two `clusters' can originate from a common ancestor, such as jet fragmentation into two $\rho$ mesons which in turn decay into two pairs of pions. In our formulism, such `clusters' are considered to be parts of a single cluster rather than two $\rho$-decay clusters.)

Again, in the special case where particle correlation in clusters does not vary with cluster location $\tphi_k$ ($\Na={\rm const.}$, $\Nhy={\rm const.}$, and $\fcl(\dphi,\tphi_k)=\fcl(\dphi)$), Eq.~(\ref{eq6}) becomes
\be
\vtwo{\acl}=\intpi\rhocl(\tphi_k)d\tphi_k\cos2\tphi_k\intpi\fcl(\dphi_i)\cphi_id\dphi_i=\vtwo{cl}\mean{\cphi}_{cl},
\label{eq8}
\ee
and we have
\be
X=\Nhy\Na\vv{2}_{hy}\vv{2}_{cl}\mean{\cphi}_{cl}.
\ee

\subsection{Particle correlation between clusters}

Correlation between particles from different clusters is given by
\bea
Y&=&\sum_{i\in k_1}\sum_{j\in k_2}\cij\nonumber\\
&=&\intpi\rhocl(\tphi_{k_1})d\tphi_{k_1}\intpi \Na(\tphi_{k_1})\fcl(\dphi_i,\tphi_{k_1})d\dphi_i\times\nonumber\\
&&\intpi\rhocl(\tphi_{k_2})d\tphi_{k_2}\intpi \Na(\tphi_{k_2})\fcl(\dphi_j,\tphi_{k_2})d\dphi_j\left[\cos2(\phi_i-\phi_j)\right],
\eea
where $k_1$ and $k_2$ stand for two clusters. Rewriting $\phi_i-\phi_j=\dphi_i+\tphi_{k_1}-\dphi_j-\tphi_{k_2}$, we obtain
\be
Y=\Na^2v^2_{2,\acl}=\Na^2\vsq{2}_{\acl},
\label{eq10}
\ee
where $\vtwo{\acl}$ is given by Eq.~(\ref{eq6}). Again the cluster particle elliptic flow squared in Eq.~(\ref{eq10}) contains flow fluctuation.

In the special case where particle correlation in clusters does not vary with cluster location $\tphi_k$, we have
\be
Y=\Na^2\vsq{2}_{cl}\mean{\cphi}^2_{cl}.
\ee

\subsection{Summary of non-flow effect from cluster correlations}

To summarize, let us now obtain the relationship between two-particle elliptic flow $\vv{2}$ that is affected by non-flow, and the real hydro-type two-particle elliptic flow $\vv{2}_{hy}$. Assuming Poisson statistics, Eq.~(\ref{eq1}) gives
\bea
N^2\vsq{2}&=&N^2_{hy}\vsq{2}_{hy}+\Ncl\Na^2\mean{\cij}_{cl}+2\Nhy\Ncl\Na\vv{2}_{hy}\vv{2}_{\acl}+\nonumber\\
&&\Ncl(\Ncl-1)\Na^2\vsq{2}_{\acl}\nonumber\\
&=&\left(\Nhy\vv{2}_{hy}+\Ncl\Na\vv{2}_{\acl}\right)^2+\Ncl\Na^2\left(\mean{\cij}_{cl}-\vsq{2}_{\acl}\right)
\eea
where $N=\Nhy+\Ncl\Na$, $\Ncl$ is average number of clusters, and we have taken distributions of total multiplicity and number of particles per cluster to be Poisson, so that $\mean{N(N-1)}=N^2$ and $\mean{\Na(\Na-1)}=\Na^2$. We have taken the number of cluster pairs to be $\Ncl(\Ncl-1)$ (i.e., not Poisson) so that the total number of pairs adds up to $N^2$. Rearranging, we have
\be
\vsq{2}=\left(\frac{\Nhy}{N}\vv{2}_{hy}+\frac{\Ncl\Na}{N}\vv{2}_{\acl}\right)^2+\frac{\Ncl\Na^2}{N^2}\left(\mean{\cij}_{cl}-\vsq{2}_{\acl}\right).
\label{eq12}
\ee
For many cluster types, Eq.~(\ref{eq12}) is generalized to
\be
\vsq{2}=\left(\frac{\Nhy}{N}\vv{2}_{hy}+\sum_{cl}\frac{\Ncl\Na}{N}\vv{2}_{\acl}\right)^2+\sum_{cl}\frac{\Ncl\Na^2}{N^2}\left(\mean{\cij}_{cl}-\vsq{2}_{\acl}\right).
\label{eq13}
\ee

We shall focus on the special case where all clusters are of the same type and particle correlation in clusters does not vary with cluster axis relative to the reaction plane. Using Eq.~(\ref{eq8}), Eq.~(\ref{eq12}) becomes
\bea
\vsq{2}&=&\left(\frac{\Nhy}{N}\vv{2}_{hy}+\frac{\Ncl\Na}{N}\vv{2}_{cl}\mean{\cphi}_{cl}\right)^2+\nonumber\\
&&\frac{\Ncl\Na^2}{N^2}\left(\mean{\cij}_{cl}-\vsq{2}_{cl}\mean{\cphi}_{cl}^2\right).
\label{eq14}
\eea
Eq.~(\ref{eq14}) can be rewritten into
\bea
\vsq{2}&=&\vsq{2}_{hy}+2\frac{\Nhy}{N}\frac{\Ncl\Na}{N}\vv{2}_{hy}\left(\vv{2}_{cl}\mean{\cphi}_{cl}-\vv{2}_{hy}\right)+\nonumber\\
&&\left(\frac{\Ncl\Na}{N}\right)^2\left(\vsq{2}_{cl}\mean{\cphi}_{cl}^2-\vsq{2}_{hy}\right)+\nonumber\\
&&\frac{\Ncl\Na^2}{N^2}\left(\mean{\cij}_{cl}-\vsq{2}_{cl}\mean{\cphi}_{cl}^2\right).
\label{eq15}
\eea
The second term on the r.h.s.~is non-flow (beyond hydro-flow) due to correlation between hydro-particles and cluster particles in excess of that between two hydro-particles, and the third term is that due to correlation between particles from different clusters. These non-flow contributions, which are beyond hydro-flow, can be positive or negative, depending on the relative magnitudes of background particle flow and cluster flow diluted by particle spread inside cluster. The non-flow contributions are positive when $\vv{2}_{cl}\mean{\cphi}_{cl}>\vv{2}_{hy}$ and negative when $\vv{2}_{cl}\mean{\cphi}_{cl}<\vv{2}_{hy}$. This can be easily understood because if $\vv{2}_{cl}=\vv{2}_{hy}$ , then the angular smearing of particles inside each cluster, $\mean{\cphi}_{cl}$, makes the angular variation of cluster particles less than that of hydro-particles, resulting in a negative non-flow contribution. If the net effect of cluster anisotropy and particle distribution inside clusters, $\vv{2}_{cl}\mean{\cphi}_{cl}$, equals to hydro anisotropy, then cluster particles and hydro-particles have the same angular variation relative to the reaction plane, resulting in zero non-flow from cross-pairs between hydro-particles and cluster particles and between particles from different clusters.

The second part of the last term of Eq.~(\ref{eq15}) r.h.s., $\frac{\Ncl\Na^2}{N^2}\vsq{2}_{cl}\mean{\cphi}_{cl}^2$, arises from the fact that number of clusters is fixed in order to have Poisson statistics for hydro particle multiplicity, particle multiplicity in clusters, and total number of particles~\cite{FWang}. It can be safely neglected because generally $\vsq{2}_{cl}<<1$. The first part of the last term of Eq.~(\ref{eq15}) r.h.s., $\frac{\Ncl\Na^2}{N^2}\mean{\cij}_{cl}$, is non-flow due to correlation between particles in the same cluster. This non-flow contribution can also be positive or negative. If particle emissions within clusters are independent, $\mean{\cij}_{cl}=\mean{\cphi}_{cl}^2$ , then Eq.~(\ref{eq15}) becomes
\bea
\vsq{2}&=&\vsq{2}_{hy}+2\frac{\Nhy}{N}\frac{\Ncl\Na}{N}\vv{2}_{hy}\left(\vv{2}_{cl}\mean{\cphi}_{cl}-\vv{2}_{hy}\right)+\nonumber\\
&&\left(\frac{\Ncl\Na}{N}\right)^2\left(\vsq{2}_{cl}\mean{\cphi}_{cl}^2-\vsq{2}_{hy}\right)+\nonumber\\
&&\frac{\Ncl\Na^2}{N^2}\left(1-\vsq{2}_{cl}\right)\mean{\cphi}_{cl}^2.
\label{eq16}
\eea
In this case the non-flow contribution due to particle correlation within clusters %(the last term in Eq.~(\ref{eq16}) r.h.s.) 
can only be positive.  

We note that the non-flow contributions from the second and third term of Eq.~(\ref{eq15}) r.h.s.~have the identical azimuthal shape relative to the reaction plane as that of hydro-flow, because they arise from the common correlation of clusters and hydro-particles to the reaction plane. As a result these non-flow contributions will unlikely be separated from medium hydro-flow in inclusive measurement of azimuthal correlation. To separate these two contributions, one needs to identify clusters and measure two-cluster azimuthal correlation. In fact, elliptic flow is often defined as the second harmonic of particle distribution relative to the reaction plane, $\vv{RP}=\mean{\cos2(\phi-\psiRP)}$ . For events composed of hydro-particles and clusters, we have
\bea
\vv{RP}&=&\frac{N_{hy}}{N}\mean{\cos2(\phi-\psiRP)}_{hy}+\frac{\Ncl\Na}{N}\mean{\cos2(\phi-\psiRP)}_{cl}\nonumber\\
&=&\frac{N_{hy}}{N}\vv{RP}_{hy}+\frac{\Ncl\Na}{N}\vv{RP}_{cl}\mean{\cphi}_{cl}.
\label{eq17}
\eea
This is analogous to the terms in the first pair of parentheses on Eq.~(\ref{eq14}) r.h.s.~except the latter contains flow fluctuation. The elliptic flow definition by Eq.~(\ref{eq17}) contains cluster contribution through angular spread of particles in clusters, $\mean{\cphi}_{cl}$, and anisotropy of the clusters themselves, $\vv{RP}_{cl}$. This raises question to comparisons often made between elliptic flow measurements and hydro calculations which may include flow fluctuation but does not include cluster correlations. 

\section{Elliptic flow for jet-correlation background}

In this section, we derive an analytical form for flow-background to jet-correlation in the cluster approach, as used in our non-flow study above and in~\cite{FWang}. We suppose a relativistic heavy-ion collision event is composed of hydrodynamic medium particles, jet-correlated particles, and particles correlated via clusters. Hydro-particles, high $\pt$ trigger particles, and clusters are distributed relative to the reaction plane ($\psi$) by
\be
\frac{dN}{d\phi}=\frac{N}{2\pi}\left[1+2v_2\cos2(\phi-\psi)\right]
\label{eqn1}
\ee	 
with the corresponding elliptic flow parameter $v_2$ and multiplicity $N$. Particle azimuthal distribution with respect to a trigger particle is
\be
\frac{1}{\Nt}\frac{dN}{d\dphi}=\frac{d\Nhy}{d\dphi}+\sum_{k\neq jet\in clus}\frac{d\Nak}{d\dphi}+\sum_{k\in jet}\frac{d\Nak}{d\dphi}+\frac{d\Njet}{d\dphi}
\label{eqn2}
\ee	 
where $\dphi=\phi-\phit$. In Eq.~(\ref{eqn2}), $d\Njet/d\dphi$ is jet-correlation signal of interest. All other terms are backgrounds. If trigger particle multiplicity is Poisson and effects due to interplay between collision centrality selection (usually via multiplicity) and trigger bias are negligible, then the background event of a triggered (di-)jet should be identical to any inclusive event, without requiring a high $\pt$ trigger particle, but with all other event selection requirements as same as for triggered events~\cite{3part_method}. Thus, we can use inclusive events %(minimum bias events of the same centrality as the triggered events) 
to obtain flow-background:
\be
\frac{1}{\Nt}\frac{dN}{d\dphi}=a\left(\frac{d\Nhy}{d\dphi}+\sum_{k\neq jet\in clus}\frac{d\Nak}{d\dphi}+\sum_{k\in jet}\frac{d\Nak}{d\dphi}\right)_{inc}+\frac{d\Njet}{d\dphi}
\ee
where $a$ is a normalization factor, often determined by the assumption of ZYAM or ZYA1 (zero jet-correlated yield at minimum or at $\dphi=1$)~\cite{jetspec}, and is approximately unity. The background is
\be
\frac{d\Nbg}{d\dphi}=\frac{d\Nhy}{d\dphi}+\sum_{k\neq jet\in clus}\frac{d\Nak}{d\dphi}+\sum_{k\in jet}\frac{d\Nak}{d\dphi}=\frac{d\Nhy}{d\dphi}+\sum_{cl}\Ncl\frac{d\Na}{d\dphi}
\ee
where we have eliminated subscript `$inc$' to lighten notation. We have summed over all cluster types `$cl$' including jet-correlation, where $\Ncl$ is number of clusters of type `$cl$'. Different cluster types include jet and minijet correlations, resonance decays, etc.

The hydro-background is simply
\be
\frac{d\Nhy}{d\dphi}=\frac{\Nhy}{2\pi}\left(1+2\vtwo{t}\vtwo{hy}\cphi\right)
\label{eqn5}
\ee
where $\vtwo{t}$ is elliptic flow parameter of trigger particles and $\vtwo{hy}$  is that of hydro-medium particles. 

The cluster particles background is given by
\be
\frac{d\Na}{d\dphi}=\intpi d\tphi_t\rho_t(\tphi_t)\intpi d\tphi_k\rhocl(\tphi_k)\intpi d\dphi_i\fcl(\dphi_i,\tphi_k)\times\frac{1}{2\pi}\delta(\dphi_i+\tphi_k-\dphi-\tphi_t)
\label{eqn6}
\ee
where $\tphi_t=\phit-\psi$, $\tphi_k=\phi_k-\psi$, $\dphi_i=\phi_i-\phi_k$, and $\rho_t(\tphi_t)=\frac{1}{2\pi}\left(1+2v_{2,t}\cos2\tphi_t\right)$ and $\rhocl(\tphi_k)=\frac{1}{2\pi}\left(1+2v_{2,cl}\cos2\tphi_k\right)$ are density profiles (i.e., $v_2$-modulated distributions) of trigger particles and clusters relative to the reaction plane, respectively. We have assumed that the cluster axis (or cluster parent) distribution is also anisotropic with respect to the reaction plane. In Eq.~(\ref{eqn6}), $\fcl(\dphi_i,\tphi_k)=\frac{d\Nak}{d\dphi_i}$ is distribution of particles in cluster relative to cluster axis (cluster correlation function), which may depend on the cluster axis relative to the reaction plane $\tphi_k$~\cite{Aoqi}. Decomposing $\rho_t(\tphi_t)$, we obtain
\bea
\frac{d\Na}{d\dphi}&=&\frac{1}{2\pi}\intpi d\tphi_k\rhocl(\tphi_k)\intpi d\dphi_i\fcl(d\dphi_i,\tphi_k)+\nonumber\\
&&\frac{2\vtwo{t}}{2\pi}\intpi d\tphi_k\rhocl(\tphi_k)\intpi d\dphi_i\fcl(\dphi_i,\tphi_k)\cos2(\dphi_i+\tphi_k-\dphi).
\eea
Because of symmetry, $\fcl(\dphi_i,\tphi_k)=\fcl(-\dphi_i,-\tphi_k)$ and $\rhocl(\tphi_k)=\rhocl(-\tphi_k)$, we have 
\be
\intpi d\tphi_k\rhocl(\tphi_k)\intpi d\dphi_i\fcl(\dphi_i,\tphi_k)\sin2(\dphi_i+\tphi_k)=0.
\ee
Therefore
\bea
\frac{d\Na}{d\dphi}&=&\frac{1}{2\pi}\intpi d\tphi_k\rhocl(\tphi_k)\Na(\tphi_k)+\nonumber\\
&&\frac{2\vtwo{t}}{2\pi}\cphi\intpi d\tphi_k\rhocl(\tphi_k)\intpi d\dphi_i\fcl(\dphi_i,\tphi_k)\cos2(\dphi_i+\tphi_k).
\label{eqn6p5}
\eea
Realizing that elliptic flow parameter of particles from clusters is given by Eq.~(\ref{eq6}),
%\be
%\vtwo{\acl}\equiv\mean{\phi-\psi}_{cl}=\frac{1}{\Na}\intpi d\tphi_k\rhocl(\tphi_k)\intpi d\dphi_k\fcl(\dphi_k,\tphi_k)\cos2(\dphi_k+\tphi_k),
%\ee
we rewrite Eq.~(\ref{eqn6p5}) into
\be
\frac{d\Na}{d\dphi}=\frac{\Na}{2\pi}\left(1+2\vtwo{t}\vtwo{\acl}\cphi\right).
\label{eqn7}
\ee

From Eq.~(\ref{eqn5}) and (\ref{eqn7}) we obtain the total background as given by
\be
\frac{d\Nbg}{d\dphi}=\frac{\Nbg}{2\pi}\left[1+2\vtwo{t}\left(\frac{\Nhy}{\Nbg}\vtwo{hy}+\sum_{cl}\frac{\Ncl\Na}{\Nbg}\vtwo{\acl}\right)\cphi\right],
\label{eqn8}
\ee
where 
\be
\Nbg=\Nhy+\sum_{cl}\Ncl\Na.
\ee
The $v_2$'s in Eqs.~(\ref{eqn5}), (\ref{eqn7}), and (\ref{eqn8}) include fluctuations, so they should be replaced by $\sqrt{\mean{v_2^2}}$. The hydro-particles $\sqrt{\mean{v_2^2}}$ is equivalent to two-particle $\vv{2}$ because there is no non-flow effect between hydro-particle pairs; same for the cluster $\sqrt{\mean{v_2^2}}$ because there is no non-flow effect between different clusters (we consider sub-clusters to be part of their parent cluster). Thus Eq.~(\ref{eqn8}) should be
\be
\frac{\Nbg}{d\dphi}=\frac{\Nbg}{2\pi}\left(1+2\vtwo{t}\vtwo{bg}\cphi\right)
\ee
where
\be
\vtwo{bg}=\frac{\Nhy}{\Nbg}\vv{2}_{hy}+\sum_{cl}\frac{\Ncl\Na}{\Nbg}\vv{2}_{\acl}.
\label{eqn11}
\ee
We note that here cluster includes single-particle (within a give $\pt$ range) cluster. Those single-particle clusters do not contribute to non-flow in $\vv{2}_{\acl}$, but they differ from single hydro-particles because they may possess different $v_2$ values.

In principle, $\vtwo{t}$ should have a similar expression as Eq.~(\ref{eqn11}) out of symmetry reason:
\be
\vtwo{t}=\frac{N_{t,hy}}{N_{t,tot}}\vv{2}_{t,hy}+\sum_{cl\_t}\frac{N_{cl\_t}N_{t,cl\_t}}{N_{t,tot}}\vv{2}_{t,cl\_t}.
\ee
where $N_{t,hy}$ is number of high $\pt$ trigger particles from hydro-medium (i.e., background trigger particles), $\vv{2}_{t,hy}$ is the elliptic anisotropy of those background trigger particles, $N_{cl\_t}$ is number of clusters of type `$cl\_t$' containing at least one trigger particle, $N_{t,cl\_t}$ is number of trigger particles per cluster, $\vv{2}_{t,cl\_t}$ is elliptic flow parameter of trigger particles from clusters, and $N_{t,tot}=N_{t,hy}+\displaystyle{\sum_{cl\_t}}N_{cl\_t}N_{t,cl\_t}$. The only difference is that trigger particles are dominated by clusters (mostly jets), and those clusters are dominated by single-trigger-particle clusters; hydro-medium contribution to trigger particle population should be small. We note that jet-correlation functions are usually normalized by total number of trigger particles including those from hydro-medium background.

If particle correlation in clusters does not vary with cluster axis relative to the reaction plane, elliptic flow of particles from clusters is given by Eq.~(\ref{eq8}), or
\be
\vv{2}_{\acl}\equiv\vv{2}_{cl}\mean{\cphi}_{cl}.
\ee
Therefore
\be 
\vtwo{bg}=\frac{\Nhy}{\Nbg}\vv{2}_{hy}+\sum_{cl}\frac{\Ncl\Na}{\Nbg}\vv{2}_{cl}\mean{\cphi}_{cl}.
\label{eqn13}
\ee

%%%%%%%%%%%%%%%%%%%%%%%%%%%%%%%%%%%%%%%%%%%%%%%%%%%%%%%%%%%%%%%%%%%%%%
\section{Jet-background $v_2$ is the reaction plane $v_2$}

Obviously, the elliptic flow in Eq.~(\ref{eqn11}) or (\ref{eqn13}) contains not only the two-particle anisotropy relative to the reaction plane, but also non-flow related to angular spread of clusters. How to obtain the elliptic flow as in Eq.~(\ref{eqn11}) or (\ref{eqn13})? In Section~\ref{sec:nonflow}, we have derived Eq.~(\ref{eq13}) for two-particle $\vv{2}$ in the cluster approach.
The quantity in the first pair of parentheses in r.h.s.~of Eq.~(\ref{eq13}) is elliptic flow due to correlation with respect to the reaction plane. The second term in the r.h.s.~arises from cluster correlation. Since elliptic flow is formally defined to be relative to the reaction plane, the first term in r.h.s.~of Eq.~(\ref{eq13}) may be considered as ``true'' elliptic flow (except flow fluctuation effect), $\vtwo{{\rm flow}}$. %We note, however, it is not necessarily as same as hydro-flow because of contamination from clusters due to coupling between cluster correlation and cluster flow. 
The second term in r.h.s.~of Eq.~(\ref{eq13}) can be considered as non-flow, $\vtwo{{\rm non-flow}}$; non-flow is due to correlations between particles from the same dijet or the same cluster. Eq.~(\ref{eq13}) can be expressed into
\be
\vsq{2}=v^2_{2,{\rm flow}}+v^2_{2,{\rm non-flow}}.
\ee

Comparing Eq.~(\ref{eq13}) with Eq.~(\ref{eqn11}), we see that 
\be
\vtwo{bg}=\vtwo{{\rm flow}},
\ee
i.e., the quantity in the first pair of parentheses in r.h.s.~of Eq.~(\ref{eq13}) is the $v_2$ parameter in Eq.~(\ref{eqn11}) that is needed in constructing jet-correlation background. In other words, elliptic flow parameter that should be used in jet-correlation flow background is the ``true'' two-particle elliptic flow (i.e., due to the reaction plane and including fluctuation).

%%%%%%%%%%%%%%%%%%%%%%%%%%%%%%%%%%%%%%%%%%%%%%%%%%%%%%%%%%%%%%%%%%%%%%
\section{Monte Carlo checks}

In this section, we verify our analytical results by Monte Carlo simulations. We generate events consisting of three components. One component is hydro-medium particles according to Eq.~(\ref{eqn1}), given hydro-particles elliptic flow parameter $\vtwo{hy}$ and Poisson distributed number of hydro-particles with average multiplicity $\Nhy$. The second component is clusters, given cluster elliptic flow parameter $\vtwo{cl}$ and fixed number of clusters $\Ncl$; each cluster is made of particles with Poisson multiplicity distribution with average $\Na$ and Gaussian azimuth spread around cluster axis with $\sigma_{\acl}$. The third component is trigger particles with accompanying associated particles; the trigger particle multiplicity is Poisson with average $\Nt$, and the elliptic flow parameter is $\vtwo{t}$. The associated particles are generated for each trigger particle by correlation function:
\bea
f(\dphi,\tphi_t)&=&C(\tphi_t)+\frac{N_{\ns}(\tphi_t)}{\sqrt{2\pi}\sigma_{\ns}(\tphi_t)}\exp\left[-\frac{(\dphi)^2}{2\sigma_{\ns}^2(\tphi_t)}\right]+\nonumber\\
&&\frac{N_{\as}(\tphi_t)}{\sqrt{2\pi}\sigma_{\as}(\tphi_t)}\left(\exp\left[-\frac{\left(\dphi-\pi+\theta(\tphi_t)\right)^2}{2\sigma_{\as}^2(\tphi_t)}\right]+\exp\left[-\frac{\left(\dphi-\pi-\theta(\tphi_t)\right)^2}{2\sigma_{\as}^2(\tphi_t)}\right]\right),
\label{eqn16}
\eea
where the near- and away-side associated particle multiplicities are Poisson with averages $N_{\ns}(\tphi_t)$ and $N_{\as}(\tphi_t)$, respectively. The Gaussian widths of the near- and away-side peaks are fixed, and the two away-side symmetric peaks are set equal and their separation is fixed. All parameters in the jet-correlation function of Eq.~(\ref{eqn16}) can be dependent on the trigger particle azimuth relative to the reaction plane, $\tphi_t$.

We first verify Eq.~(\ref{eq13}) by generating events with hydro-particles and jet-correlated particles. (We do not include other clusters except jet-correlations.) We use $\Nhy=150$, $\vtwo{hy}=0.05$, and $\Nt=2$, $\vtwo{t}=0.5$. We use the large trigger particle $v_2$ in order to maximize the effect of non-flow. For jet-correlation function, we generate back-to-back dijet with $N_{\ns}=0.7$, $N_{\as}=1.2$, $\sigma_{\ns}=0.4$, $\sigma_{\as}=0.7$, and $\theta=0$ (referred to as dijet model). We fix $v_2$ in the simulation, i.e., $v_2$ fluctuation is not included. We simulate $10^6$ events and calculate $\vsq{2}=\mean{\cij}$. Including only hydro-particles, we obtain $\vv{2}_{hy}=0.05005\pm0.00009$, consistent with the input. Including all simulated events and all particles (hydro-particles and jet-correlated particles), we obtain $\vv{2}_{inc}=0.05560\pm0.00008$. Using triggered events (events containing at least one trigger particle) only, we obtain $\vv{2}_{trig\_evt}=0.05642\pm0.00008$. Using triggered events but excluding one dijet at a time (i.e., using the underlying background event of the dijet) and repeating over all dijets in the event, we obtain $\vv{2}_{bg}=0.05568\pm0.00008$. We see that the background $v_2$ is as same as that obtained from inclusive events, $\vv{2}_{bg}=\vv{2}_{inc}$, and both are smaller than that from triggered events only.

We can in fact predict the inclusive event $v_2$ by Eq.~(\ref{eqn13}) using the ``hydro + dijet'' model. The average $\sqrt{\mean{\cij}}$ of jet-correlated particle pairs within the same dijet is $\sqrt{\mean{\cij}}_{jet}=\mean{\cphi}_{jet}=0.5054\pm0.0004$. This is consistent with the expected value 
\[\mean{\cphi}_{jet}=\frac{N_{\ns}}{N_{\ns}+N_{\as}}\exp\left(-2\sigma_{\ns}^2\right)+\frac{N_{\as}}{N_{\ns}+N_{\as}}\exp\left(-2\sigma_{\as}^2\right)\cos2\theta=0.5046\]
where $\theta=0$.
The average $\sqrt{\mean{\cij}}$ of pairs of particles from different dijets is $0.2516\pm0.0008$; it equals to $\vv{2}_{a,jet}=\vtwo{t}\mean{\cphi}_{jet}=0.5\times0.5046=0.2523$. The average $\sqrt{\mean{\cij}}$ for cross-talk pairs of background particle and jet-correlated particle is $0.10064\pm0.00004$; and it equals to the expected value $\sqrt{\vv{2}_{hy}\vtwo{t}\mean{\cphi}_{jet}}=\sqrt{0.05\times0.5\times0.5046}=0.1123$. The inclusive event two-particle elliptic flow parameter is 
\[\vv{2}=\sqrt{\left(\frac{150}{153.8}\times0.05+\frac{3.8}{153.8}\times0.2523\right)^2+\frac{2\times1.9^2}{153.8^2}\left(0.5054^2-0.2523^2\right)}=%\sqrt{0.05500^2+0.00765^2}=
0.05553;\]
this is indeed consistent with $\vv{2}_{inc}$ or $\vv{2}_{bg}$ obtained from simulation.

We now verify Eq.~(\ref{eqn11}) or (\ref{eqn13}) as the correct $v_2$ to be used for jet-correlation background subtraction. We generate Poisson distributed hydro-particles with average multiplicity $\Nhy=150$ and fixed elliptic flow parameter $\vtwo{hy}=0.05$. We generate Poisson distributed trigger particles with average trigger particle multiplicity $\Nt=2.0$; we use different jet-correlation functions (discussed below). We also include clusters that do not have trigger particles (referred to as minijet clusters); the particle multiplicity per minijet cluster is Poisson distributed with average $\Na=5$, and the number of minijet clusters is fixed as $\Ncl$; we fix the cluster shape to be Gaussian with width $\sigma_{\acl}=0.5$ (and average angular spread $\mean{\cphi}_{cl}=\exp\left(-2\sigma^2_{\acl}\right)=0.6065)$, and also fix the cluster elliptic flow parameter $\vtwo{cl}=0.20$. We simulate $10^6$ events and form raw correlation functions normalized by the number of trigger particles. In order to extract the real background $v_2$ from the simulations, we subtract the input jet-correlation function. If the jet-correlation function varies with the trigger particle angle relative to the reaction plane, the trigger multiplicity weighted average jet-correlation function is subtracted. We fit the resultant background function to $B\left(1+2\vtwo{t}\vtwo{{\rm fit}}\cphi\right)$ where $B$ and $\vtwo{{\rm fit}}$ are fit parameters. We treat the input $\vtwo{t}$ as known; we did not include any complication into $\vtwo{t}$. We compare the fit $\vtwo{{\rm fit}}$ to the calculated one by Eq.~(\ref{eqn11}) or (\ref{eqn13}). We study several cases with different shapes for jet-correlation function, as well as varying values for some of the input parameters: 
\begin{itemize}
\item[(i)] ``hydro + dijet'' model: we generate back-to-back dijets accompanying trigger particles, without other clusters. The calculated $\vtwo{bg}$ by Eq.~(\ref{eqn13}) is $\vtwo{bg}=\frac{150}{153.8}\times0.05+\frac{2\times1.9}{153.8}\times0.5\times0.5046=0.05500$. 
\item[(ii)] ``hydro + minijet + dijet'' model: we include minijet clusters in addition to (i). The calculated $\vtwo{bg}$ by Eq.~(\ref{eqn13}) is $\vtwo{bg}=\frac{150}{203.8}\times0.05+\frac{5\times10}{203.8}\times0.2\times0.6065+\frac{2\times1.9}{203.8}\times0.5\times0.5046=0.07126$. 
\item[(iii)] ``hydro + minijet + near-side + away-side double-peak'' model: we generate jet-correlated particles by correlation function with double-peak away-side to replicate the experimentally measured reaction-plane averaged dihadron correlation function~\cite{Horner,Jia}. We used the same Gaussian parameters for the correlation peaks as in (i) but $\theta=1$, thus $\mean{\cphi}_{jet}=0.1689$. The calculated $\vtwo{bg}$ by Eq.~(\ref{eqn13}) is $\vtwo{bg}=\frac{150}{203.8}\times0.05+\frac{5\times10}{203.8}\times0.2\times0.6065+\frac{2\times1.9}{203.8}\times0.5\times0.1689=0.06813$. 
\item[(iv)] ``hydro + minijet + near-side + reaction-plane dependent away-side double-peak'' model: we include reaction-plane dependent jet-correlation function similar to preliminary experimental data~\cite{Aoqi}. We have to use Eq.~(\ref{eqn11}) to calculate $\vtwo{bg}$, which gives $\vtwo{bg}=\frac{150}{203.5}\times0.05+\frac{5\times10}{203.5}\times0.2\times0.6065+\frac{2\times1.74}{203.5}\times0.1423=0.06910$. Note that, in this simulation of reaction-plane dependent jet-correlation signal, the number of jet-correlated particles is not 1.9, but rather 1.74. Also note that, due to the reaction-plane dependency of the jet-correlation signal, the elliptic anisotropy of jet-correlated particles cannot be factorized into the product of the trigger particle elliptic flow and the average angular spread of the jet-correlation signal as in Eq.~\ref{eq8}, but has to be calculated by Eq.~\ref{eq6}.
\end{itemize}

We list our comparison in Table~\ref{tab}. The fit $\vtwo{{\rm fit}}$ is supposed to be the real background $\vtwo{bg}$. The fit errors are due to statistical fluctuations in the simulation. As can be seen, the calculated $\vtwo{bg}$ reproduces the real background $\vtwo{bg}$ in every case. The $\vtwo{bg}$ values differ from the hydro-background $v_2$ due to contributions from cross-talks between cluster correlation and cluster flow. %jet-correlated particles and hydro-particles and between cluster particles and hydro-particles. 
Also shown in Table~\ref{tab} are the two-particle $\vv{2}$ from all pairs in inclusive events. The $\vv{2}$ values differ from $\vtwo{bg}$ due to %additional 
non-flow contributions between particles from the same dijet or the same cluster.

\begin{table}
\caption{Monte Carlo verification of analytical results of elliptic flow parameter to be used in jet-correlation background. Hydro-particle multiplicity, trigger particle multiplicity, jet-correlated near- and away-side multiplicities, and particle multiplicity per minijet cluster are all generated with Poisson distributions, with averages $\Nhy$, $\Nt$, $N_{\ns}$, $N_{\as}$, and $\Na$, respectively, while number of minijet clusters is fixed as $\Ncl$. The jet-correlation function is given by Eq.~(\ref{eqn16}), with near- and away-side Gaussian width fixed to be $\sigma_{\ns}=0.4$ and $\sigma_{\as}=0.7$, respectively. The minijet cluster Gaussian width is fixed to $\sigma_{\acl}=0.5$. The elliptic flow parameters for hydro-particles, trigger particles, and clusters are $\vtwo{hy}$, $\vtwo{t}$, and $\vtwo{cl}$, respectively, and are fixed over all events without fluctuation. We use $\Nhy=150$, $\Nt=2$, $\Na=5$, $\vtwo{hy}=0.05$, and $\vtwo{cl}=0.20$. Four cases are studied; the parameters for the cases are, respectively,
(i) $\Ncl=0$, $\vtwo{t}=0.5$, $C=0$, $N_{\ns}=0.7$, $N_{\as}=1.2$, $\sigma_{\ns}=0.4$, $\sigma_{\as}=0.7$, and $\theta=0$; 
(ii) $\Ncl=10$, $\vtwo{t}=0.5$, $C=0$, $N_{\ns}=0.7$, $N_{\as}=1.2$, $\sigma_{\ns}=0.4$, $\sigma_{\as}=0.7$, and $\theta=0$; 
(iii) $\Ncl=10$, $\vtwo{t}=0.5$, $C=0$, $N_{\ns}=0.7$, $N_{\as}=1.2$, $\sigma_{\ns}=0.4$, $\sigma_{\as}=0.7$, and $\theta=1$; and
(iv) $\Ncl=10$, $\vtwo{t}=0.1$, $C=0$, $N_{\ns}=0.7$, $N_{\as}=1.2$, $\sigma_{\ns}=0.4$, $\sigma_{\as}=0.7$, and $\theta=1$.
}
\label{tab}
\begin{center}
\begin{tabular}{rlccc}\hline\hline
& Case & $\vv{2}$ & $\vtwo{{\rm fit}}$ & Calculated $\vtwo{bg}$\\\hline
(i) & hydro + dijet & 0.05557(8)	& 0.05505(8)	& 0.05500\\
(ii) & hydro + minijet + dijet & 0.08465(6)	& 0.07115(8)	& 0.07126\\
(iii) & hydro + minijet + near-side + & & & \\
& away-side double-peak & 0.08172(6)	& 0.06815(8)	& 0.06813\\
(iv) & hydro + minijet + near-side + & & & \\
& reaction-plane dependent away-side & 0.08279 & 0.06883(35) &	0.06910 \\
& double-peak + clusters & & & \\\hline\hline
\end{tabular}
\end{center}
\end{table}

\begin{figure*}[hbt]
\centerline{
\includegraphics[width=0.33\textwidth]{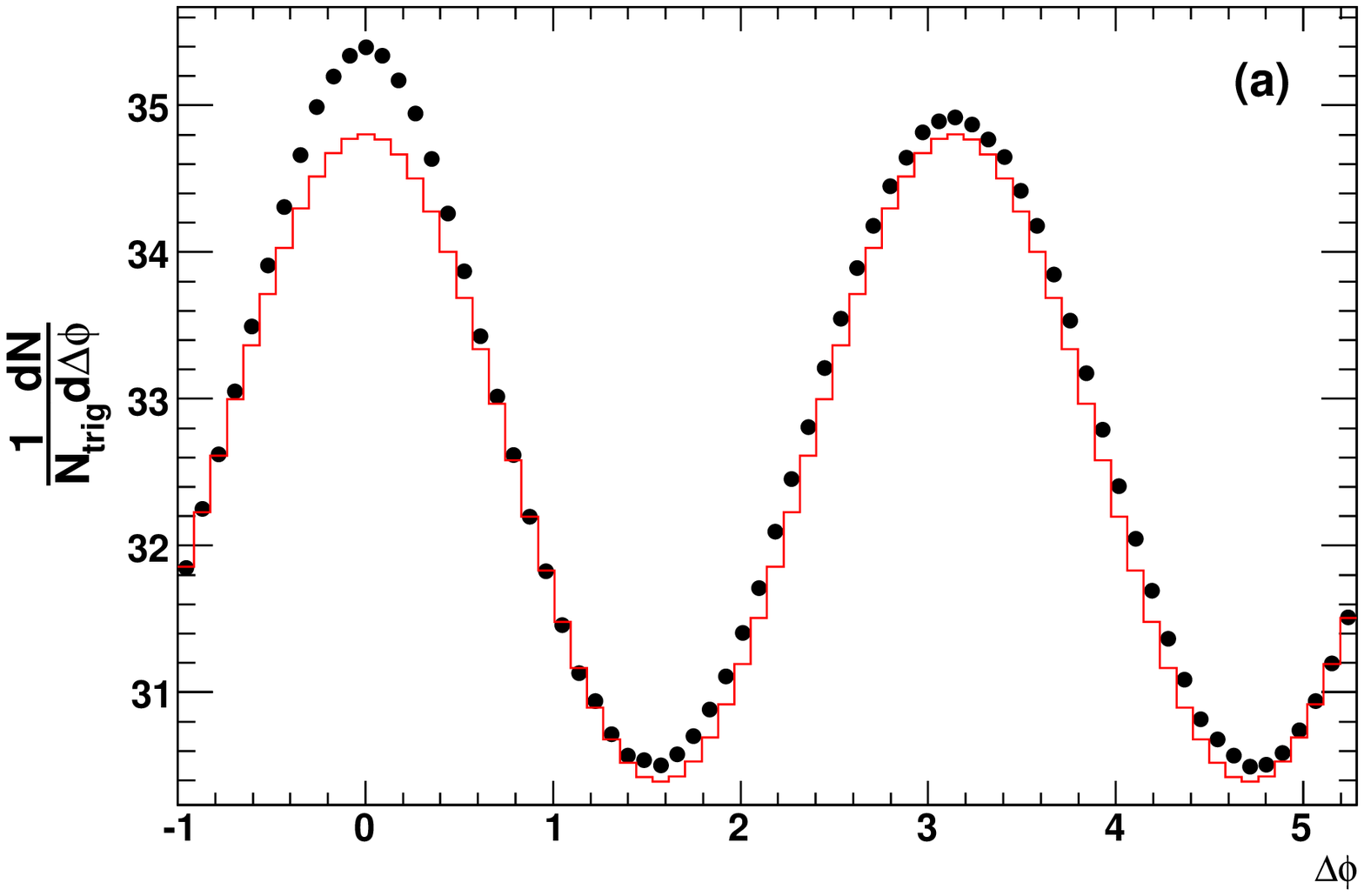}
\includegraphics[width=0.33\textwidth]{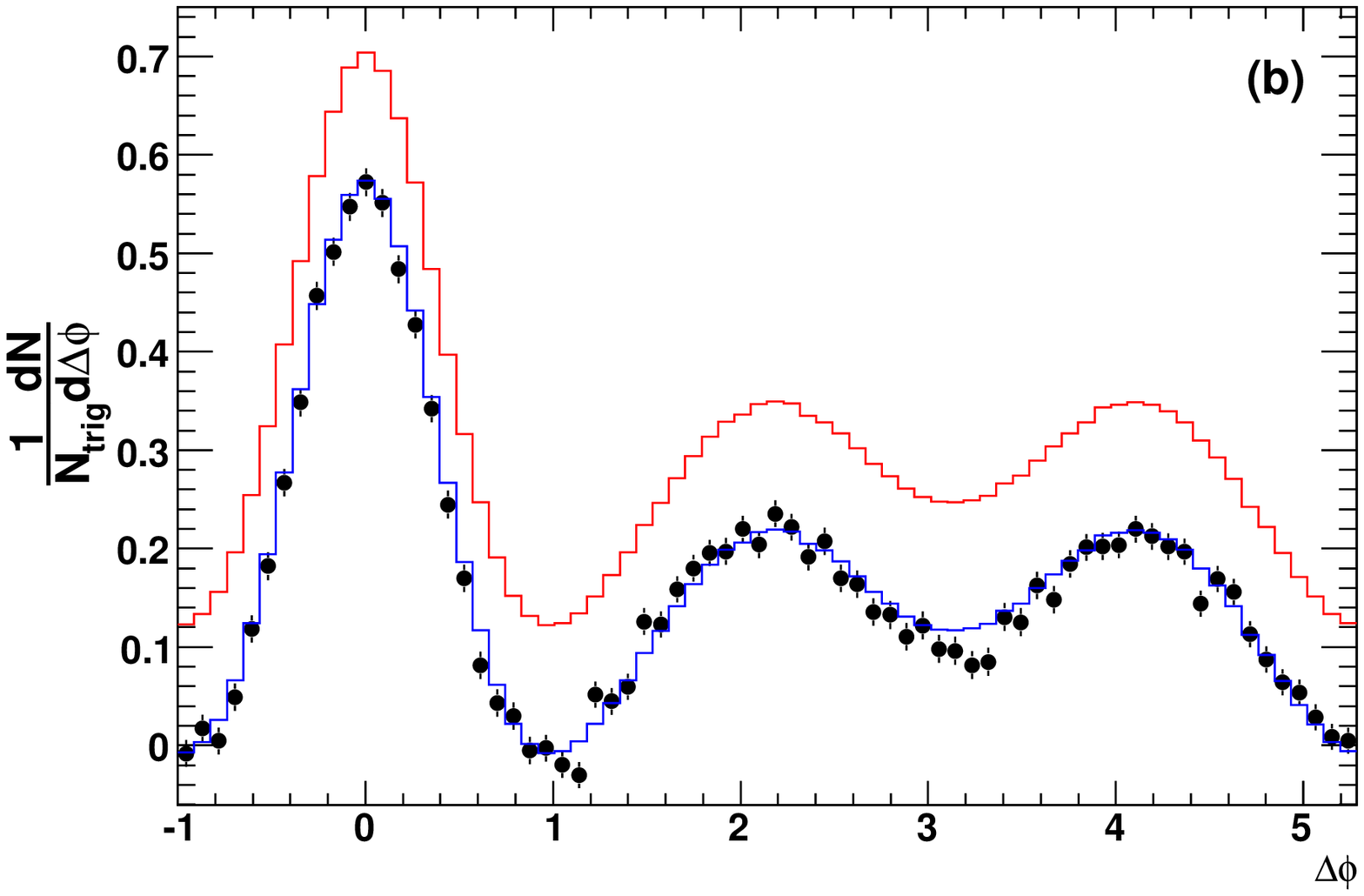}
\includegraphics[width=0.33\textwidth]{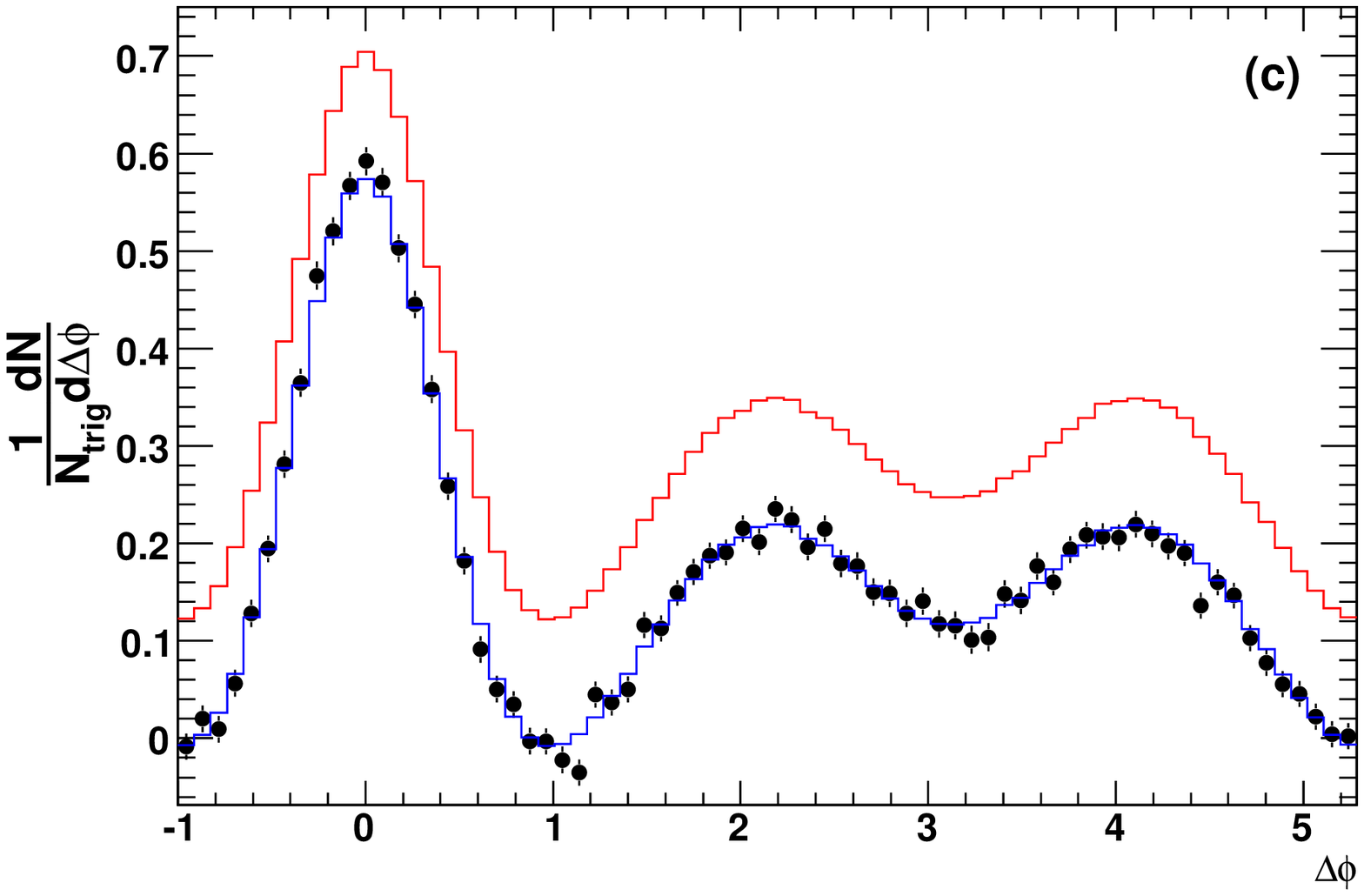}
}
\caption{(Color online) (a) Simulated raw correlation from the ``hydro + minijet + near-side + away-side double-peak'' model, Case (iii) in Table~\ref{tab}. ZYA1-normalized flow-background using the calculted $\vtwo{bg}$ is shown as the red histogram. (b) Background-subtracted jet-correlation (black data points) compared to the input correlation signal (red histogram). The background uses the calculated $\vtwo{bg}$ and is normalized to signal by ZYA1. The input signal shifted down by a constant is shown in the blue histogram. (c) As same as (b) except the subtracted background uses the decomposed $\vv{2D}$.}
\label{fig}
\end{figure*}

Figure~\ref{fig}(a) shows the raw correlation function for case (iii) and flow background using the calculated $\vtwo{bg}$ and normalized by ZYA1. Figure~\ref{fig}(b) shows the ZYA1-background subtracted jet-correlation function, using the calculated $\vtwo{bg}$ by Eq.~\ref{eqn11} for flow background. The background-subtracted jet-correlation is compared to the input signal. As shown in Fig.~\ref{fig}(b), the shapes of the input signal and extracted signal are the same, which is not surprising because the calculated $\vtwo{bg}$ is the correct value to use in flow background subtraction. The roughly constant offset is due to ZYA1-normalization.

%%%%%%%%%%%%%%%%%%%%%%%%%%%%%%%%%%%%%%%%%%%%%%%%%%%%%%%%%%%%%%%%%%%%%%
%\section{How to experimentally obtain elliptic flow for jet-correlation background}
\section{How to ``measure'' jet-correlation background (reaction plane) $v_2$}

Two-particle angular correlation is analyzed by STAR and is decomposed into two components~\cite{minijet}: one is the azimuth quadrupole, $\vv{2D}$, that is due to correlations of particles to a common source, the reaction plane; the other is minijet correlation that is due to angular correlation between particles from the same minijet or the same cluster. Properly decomposed, the azimuth quadrupole should correspond to the first term in r.h.s.~of Eq.~(\ref{eq13}),
\be
\vv{2D}=\frac{\Nhy}{\Nbg}\vv{2}_{hy}+\sum_{cl}\frac{\Ncl\Na}{\Nbg}\vv{2}_{\acl}.
\ee
%which is the elliptic flow parameter that should be used in jet-correlation background subtraction.
This is identical to Eq.~(\ref{eqn11}). That is, the elliptic flow parameter from a proper 2D quadrupole-minijet decomposition is exactly what is needed for jet-correlation background calculation. Decomposition of minijet correlation and flow, assuming the functional form for minijet correlation, has been carried out by STAR as a function of centrality but including all $\pt$~\cite{minijet}. One may restrict to narrow $\pt$ windows to obtain $\vv{2D}$ as a function of $\pt$, however, statistics can quickly run out with increasing $\pt$ because the 2D decomposition method requires particle pairs.

Figure~\ref{fig}(c) shows the ZYA1-background subtracted jet-correlation function, using the decomposed $\vv{2D}$ from the simulation data for flow background. The background-subtracted jet-correlation is compared to the input signal. The shapes of the input signal and the extracted signal are the same, which demonstrates that the decomposed $\vv{2D}$ is close to the input elliptic flow value. Again, the roughly constant offset is due to ZYA1-normalization.

One natural question to ask is why not to decompose jet-correlation and jet-background directly from high-$\pt$ triggered correlation function. One obvious reason is that jet-correlation shape is unknown a priori, thus one cannot simply fit triggered correlation to a given functional form. However, even when the functional form of jet-correlation signal is known, as is the case in our simulation, we found that the decomposed jet-correlation signal shape deviates significantly from the input one. This is because the jet-correlation signal is not orthogonal to flow background, but rather entangled, both with near- and away-side peaks, and hence one can get false minimum $\chi^2$ in decomposing the two components with limited statistics. 

%When extracting $\vv{2D}$ from correlation between any two particles (untriggered correlation), as done above, the same issue is present. However, elliptic flow background pairs dominate particle pair sample in untriggered correlation; signal correlations are negligible, particularly on the away side. Thus the extracted $\vv{2D}$ parameter from untriggered particle correlation is fairly reliable. In other words, although extraction of $\vv{2D}$ from untriggered particle correlation also involves assumption of physical model for the correlation signal, the signal has relatively weak influence on total correlation because of the large combinatorial background. We note that associated particles correlated with high-$\pt$ trigger particles constitute a very small fraction of the total sample of untriggered particle pairs, having negligible contribution to the untriggered correlation signal. Therefore, practically no entanglement is present between the extracted $\vv{2D}$ and high-$\pt$ triggered jet-correlation signal.

%%%%%%%%%%%%%%%%%%%%%%%%%%%%%%%%%%%%%%%%%%%%%%%%%%%%%%%%%%%%%%%%%%%%%%
\section{Implications and Summary}

In experimental analysis, $v_2$ values from various methods have been used for jet-correlation background. STAR used the average of the reaction plane $\vv{RP}$ and the four-particle $\vv{4}$ and used the range between them (or between $\vv{2}$ and $\vv{4}$) as systematic uncertainties~\cite{jetspec,3part}. The reaction plane $\vv{RP}$ and the two-particle $\vv{2}$ contain significant non-flow contributions, while the non-flow contributions are significantly reduced in the four particle $\vv{4}$. On the other hand, any $v_2$ fluctuation reduces $\vv{4}$. The recently measured $\vv{2D}$ magnitudes from STAR are similar to the $\vv{4}$ results, suggesting that the used $v_2$ values for background calculation are too large, by about $1\sigma$ systematic uncertainty. PHENIX used $v_2$ results from the reaction plane method where the reaction plane is determined by particles several units of pseudo-rapidity away from particles used in jet-correlation analysis. Some but not all non-flow is removed; the remaining non-flow may be dominated by the long range $\Delta\eta$ correlation (ridge) observed in non-peripheral heavy-ion collisions~\cite{jetspec,Joern,PHOBOS}. Thus the $v_2$ values used by PHENIX for background calculation are also too large.

In summary, we have derived an analytical form for jet-correlation flow-background in a cluster approach. We argue that the elliptic flow $v_2$ parameter to be used in jet-correlation background is that from two-particle method excluding non-flow correlation unrelated to the reaction plane, but including cross-terms between cluster correlation and cluster flow. We verify our result by Monte Carlo simulation for various jet-correlation signal shapes as well as varying other input parameters to the simulation. 
We demonstrate that the $v_2$ parameter to use in jet-correlation flow background is as same as the $\vv{2D}$ from a proper 2D quadrupole-minijet decomposition of two-particle angular correlation. However, we note that 2D quadrupole-minijet decomposition requires a model for minijet correlation shape, which gives rise to systematic uncertainty on the extracted $\vv{2D}$. 

%%%%%%%%%%%%%%%%%%%%%%%%%%%%%%%%%%%%%%%%%%%%%%%%%%%%%%%%%%%%%%%%%%%%%%
\section{Acknowledgment}

%We thank Dr.~Art Poskanzer, Dr.~Aihong Tang, Dr.~Tom Trainor, and Dr.~Sergei Voloshin for helpful discussions. 
F.W.~thanks Dr.~Jan Rak for invitation to the stimulating workshop.
This work was supported by U.S. Department of Energy under Grant DE-FG02-88ER40412.

%%%%%%%%%%%%%%%%%%%%%%%%%%%%%%%%%%%%%%%%%%%%%%%%%%%%%%%%%%%%%%%%%%%%%%


\begin{thebibliography}{99}
%\bibitem{flow} K.~H.~Ackermann {\it et al.} (STAR Collaboration), Phys.~Rev.~Lett. {\bf 86}, 402 (2001).

%\bibitem{whitepaper} J.~Adams {\it et al.} (STAR Collaboration), Nucl.~Phys. {\bf A757}, 102 (2005).

%\bibitem{nonflow} J.~Adams {\it et al.} (STAR Collaboration), Phys.~Rev.~Lett. {\bf 93}, 252301 (2004).

%\bibitem{v2method} A.~M.~Poskanzer and S.~A.~Voloshin, Phys.~Rev.~C {\bf 58}, 1671 (1998).

%\bibitem{v4} C.~Adler {\it et al.} (STAR Collaboration), Phys.~Rev.~C {\bf 66}, 034904 (2002).

%\bibitem{note} We have examined non-uniform decay angle distributions due to possible polarization effect, and found they do not change our qualitative results. 

%\bibitem{rho} J.~Adams {\it et al.} (STAR Collaboration), Phys.~Rev.~Lett. 92, 092301 (2004).

%\bibitem{v2data} J.~Adams {\it et al.} (STAR Collaboration), Phys.~Rev.~C {\bf 72}, 014904 (2005).

%\bibitem{spectra} J.~Adams {\it et al.} (STAR Collaboration), Phys.~Rev.~Lett. {\bf 92}, 112301 (2004).

%\bibitem{kstar} C.~Adler {\it et al.} (STAR Collaboration), Phys.~Rev.~C {\bf 66}, 061901 (2002).

%\bibitem{Levente} B.~I.~Abelev {\it et al.} (STAR Collaboration), arXiv:0808.2041.

%\bibitem{minijet} M.~Daugherity (STAR Collaboration), arXiv:0806.2121.

%\bibitem{trainor} T.~A.~Trainor, arXiv:0803.4002.

\bibitem{STARwp} J.~Adams {\it et al.} (STAR Collaboration), Nucl.~Phys.{\bf A757}, 102 (2005).

\bibitem{PHENIXwp} K.~Adcox {\it et al.} (PHENIX Collaboration), Nucl.~Phys.~{\bf A757} 184 (2005). 

\bibitem{v4} For example, C.~Adler {\it et al.} (STAR Collaboration), Phys.~Rev.~C {\bf 66}, 034904 (2002).

\bibitem{v2data} J.~Adams {\it et al.} (STAR Collaboration), Phys.~Rev.~C {\bf 72}, 014904 (2005).

\bibitem{v2method} A.~M.~Poskanzer and S.~A.~Voloshin, Phys.~Rev.~C {\bf 58}, 1671 (1998).

\bibitem{v2method2} S.~A.~Voloshin, A.~M.~Poskanzer, and R.~Snellings, arXiv:0809.2949 (2008).

\bibitem{FWang} Q.~Wang and F.~Wang, arXiv:0812.1176 (2008). 

\bibitem{FWang2} Q.~Wang and F.~Wang, arXiv:0901.0703 (2009). 

\bibitem{3part} B.~I.~Abelev {\it et al.} (STAR Collaboration), Phys.~Rev.~Lett.~{\bf 102}, 052302 (2009).

\bibitem{3part_method}
J.G.~Ulery and F.~Wang, Nucl.~Instrum.~Meth.~{\bf A595}, 502 (2008).

\bibitem{Aoqi} A.~Feng (STAR Collaboration), J.~Phys.~G {\bf 35}, 104082 (2008).

\bibitem{minijet} M.~Daugherity (STAR Collaboration), J.~Phys.~G {\bf 35}, 104090 (2008).

\bibitem{Horner} M.~J.~Horner (STAR Collaboration), J.~Phys.~G {\bf 34}, S995 (2007).

\bibitem{Jia} A.~Adare {\it el al.} (PHENIX Collaboration), Phys.~Rev.~C {\bf 78}, 014901 (2008).

\bibitem{jetspec} J.~Adams {\it et al.} (STAR Collaboration), Phys.~Rev.~Lett.~{\bf 95}, 152301 (2005).

\bibitem{Joern} J.~Putschke (STAR Collaboration), J.~Phys.~G {\bf 34}, S679 (2007). 

\bibitem{PHOBOS} E.~Wenger (PHOBOS Collaboration), J.~Phys.~G {\bf 35}, 104080 (2007). 

\end{thebibliography}
\end{document}